\documentclass[11pt]{article}
\usepackage{moriond,epsfig}

\def\Journal#1#2#3#4{{#1} {\bf #2}, #3 (#4)}

\def\NIMA{{\em Nucl. Instrum. Methods} A}

\def\PLB{{\em Phys. Lett.}  B}
\def\PRL{\em Phys. Rev. Lett.}
\def\PRC{{\em Phys. Rev.} C}
\def\PRD{{\em Phys. Rev.} D}

\def\be{\begin{equation}}
\def\ee{\end{equation}}
\def\bea{\begin{eqnarray}}
\def\eea{\end{eqnarray}}

\begin{document}
\vspace*{4cm}
\title{HIGH PRECISION MEASUREMENTS OF THE PION PROTON DIFFERENTIAL CROSS
SECTION}

\author{I.G.~ALEKSEEV, I.G.~BORDYUZHIN, D.A.~FEDIN, V.P.~KANAVETS , L.I.~KOROLEVA, 
B.V.~MOROZOV, V.M.~NESTEROV, V.V.~RYLTSOV, A.D.~SULIMOV, D.N.~SVIRIDA}

\address{Institute for Theoretical and Experimental Physics,\\
B. Cheremushkinskaya 25, Moscow, Russia}

\author{V.A.~ANDREEV, YE.A.~FILIMONOV, V.V.~GOLUBEV, A.B.~GRIDNEV, E.A.~KONOVALOVA, A.I.~KOVALEV, 
N.G.~KOZLENKO, V.S.~KOZLOV, A.G.~KRIVSHICH, D.V.~NOVINSKY, V.V.~SUMACHEV, V.I.~TARAKANOV, 
V.YU.~TRAUTMAN}

\address{Petersburg Nuclear Physics Institute,\\
Gatchina, Leningrad district, Russia}

\author{M.~SADLER}

\address{Abilene Christian University, Abilene, Texas, USA}

\maketitle\abstracts{
Study of the elastic scattering can produce a rich information on the dynamics of the strong interaction.
The EPECUR collaboration is aimed at the research of baryon resonances in the
second resonance region via pion-proton elastic scattering and kaon-lambda production.
The experiment features high statistics and better than 1~MeV resolution in the invariant mass
thus allowing searches for narrow resonances with the coupling to the $\pi p$ channel as low as 5\%.
The experiment is of "formation" type, i.e. the resonances are produced in s-channel and the scan 
over the invariant mass is done by the variation of the incident pion momentum which is measured 
with the accuracy of 0.1\% with a set of 1~mm pitch proportional chambers located in the first 
focus of the beam line. The reaction is identified by a magnetless spectrometer 
based on wire drift chambers with a hexagonal structure. Background suppression in this 
case depends on the angular resolution, so the amount of matter in the chambers and the setup was
minimized to reduce multiple scattering.  
The measurements started in 2009 with the setup optimized for elastic pion-proton scattering.
With 3 billions of triggers already recorded the differential cross section of the elastic 
$\pi p$-scattering on a liquid hydrogen target in the region of the diffraction minimum 
is measured with statistical accuracy about 1\% in 1~MeV steps in terms of the invariant mass. 
The paper covers the experimental setup, current status and some preliminary results.}

\section{Introduction}
   An interest to this experiment originated with the discovery in 2003 by the two experiments LEPS~\cite{LEPS} and
DIANA~\cite{DIANA} a new baryonic state $\theta^+$ with positive strangeness and very small width. Later 
appeared several strong results where the state was not seen~\cite{CLAS} but recent results from LEPS~\cite{LEPS1}
and DIANA~\cite{DIANA1}, as well as interference analysis of the same CLAS data~\cite{CLAS2}
still insist on the evidence for this resonance. Quantum numbers of $\theta^+$ are not
measured but it is believed that it belongs to pentaquark antidecuplet predicted in 1997 by D.~Diakonov, V.~Petrov
and M.~Polyakov~\cite{DPP}. In this case there should also exist a non-strange neutral resonance P11 with mass near 
1700~MeV. Certain hints in favour of its presence were found in the modified PWA of GWU group~\cite{GWU}
at masses 1680 and 1730~MeV~\cite{PWAM}. Recently an indication for this narrow state was found in 
$\eta$-photoproduction off neutron in GRAAL~\cite{GRAAL1,GRAAL2}.
The structure observed had a mass of 1685~MeV and a width less than $30$~MeV, which was determined by the detector resolution.
No counterpart was found in the measurements off proton. Similar peak-like structure was also observed in
experiments Tohoku-LNS~\cite{LNS}, CB-ELSA~\cite{CBELSA} and CB-MAMI~\cite{CBMAMI}. 

   Our idea is to search for P11(1700) in formation-type experiment on a pion beam~\cite{EPECUR}. 
Precise measurement of the beam momentum and fair statistics will allow us to do a scan with unprecedented
invariant mass resolution. We plan to measure differential cross sections of the reactions $\pi^-p\to\pi^-p$ and 
$\pi^-p\to K^0_S\Lambda^0$ with high statistics and better than a MeV invariant mass resolution. If the resonance 
does exist our experiment will provide statistically significant result and we will measure its width with the
precision better than 0.7~MeV.

\section{Experimental conditions}

\begin{figure}[t!]
    \centering
    \includegraphics[width=\textwidth]{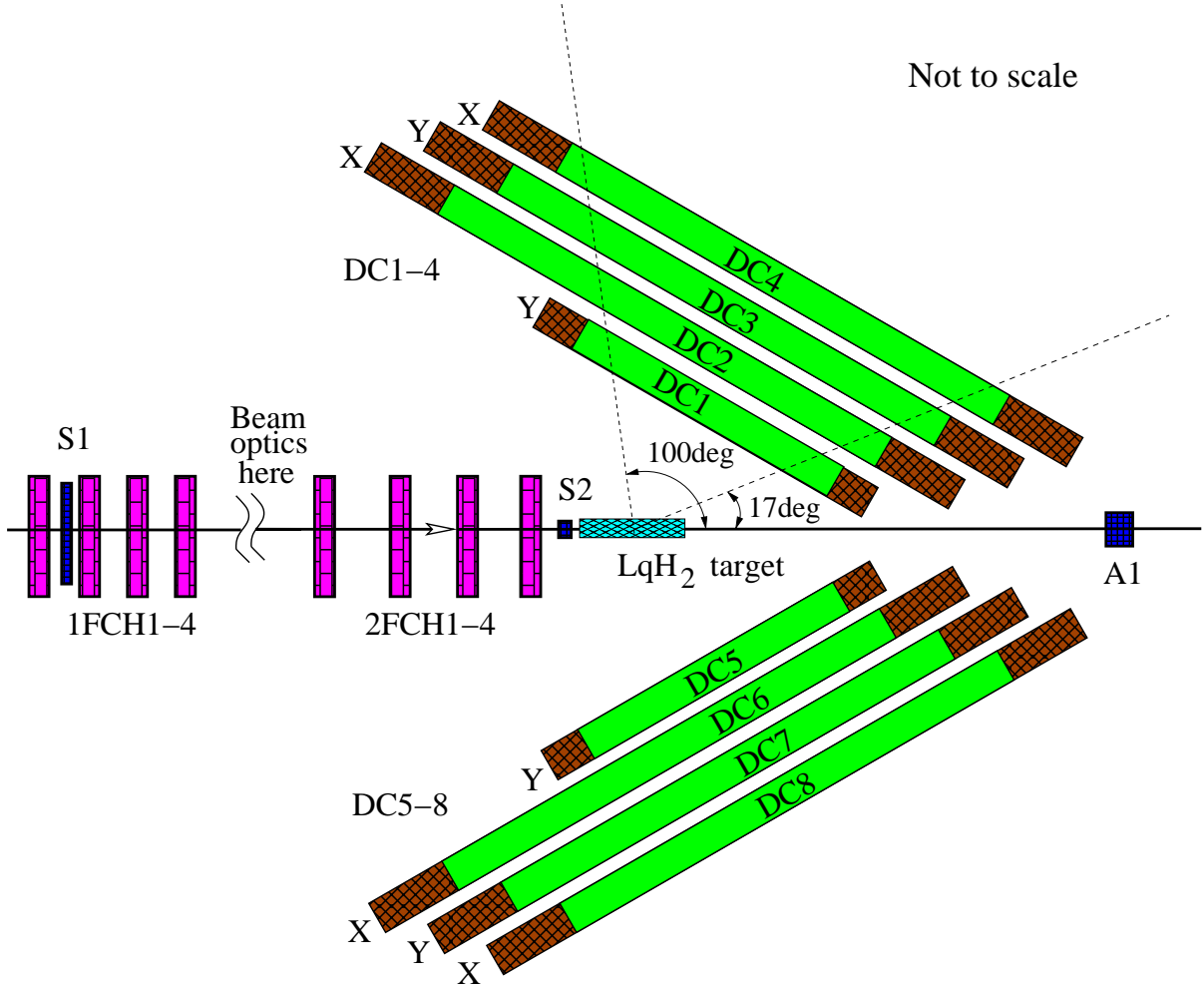}
    \caption{The experimental layout for $\pi p$ elastic scattering.}
    \label{fig:elast_setup}
\end{figure}

   The layout of the apparatus for measurements of differential cross sections of the pion-proton
elastic scattering is shown in fig.~\ref{fig:elast_setup}. The main parts are:
proportional chambers in the first (\textbf{1FCH}) and the second (\textbf{2FCH}) focuses of the
magneto-optic channel, 8 planes of drift chambers, forming the left (\textbf{DC1-4}) and the 
right (\textbf{DC1-4}) arms of the symmetric two-arm spectrometer, the liquid hydrogen target
(\textbf{LqH}$_2$) and scintillation counters (\textbf{S1}, \textbf{S2} and \textbf{A1}).

\begin{figure}[t!]
    \centering
    \begin{tabular}{cc}
    \includegraphics[width=0.45\textwidth]{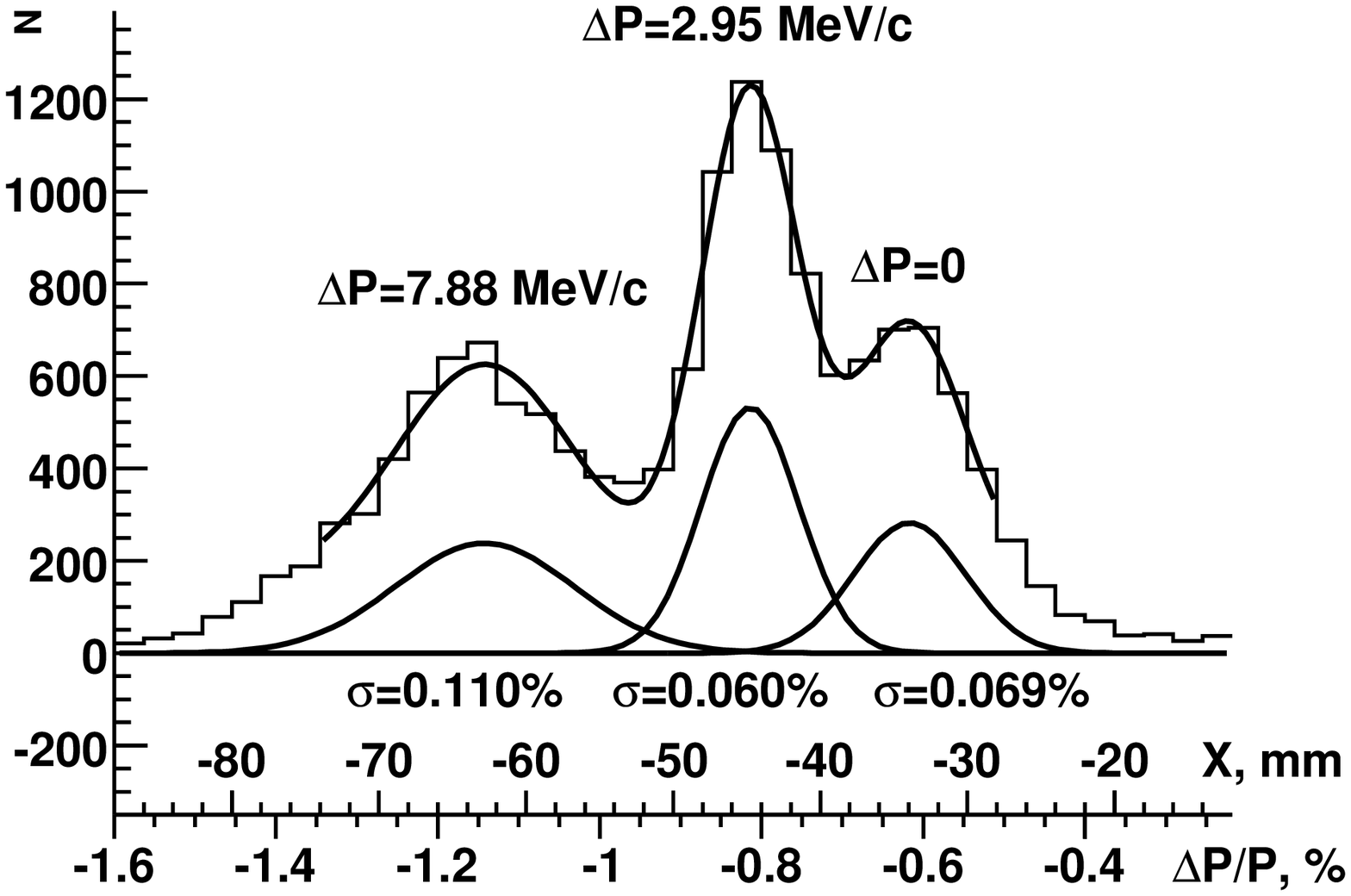} &  
    \includegraphics[width=0.45\textwidth]{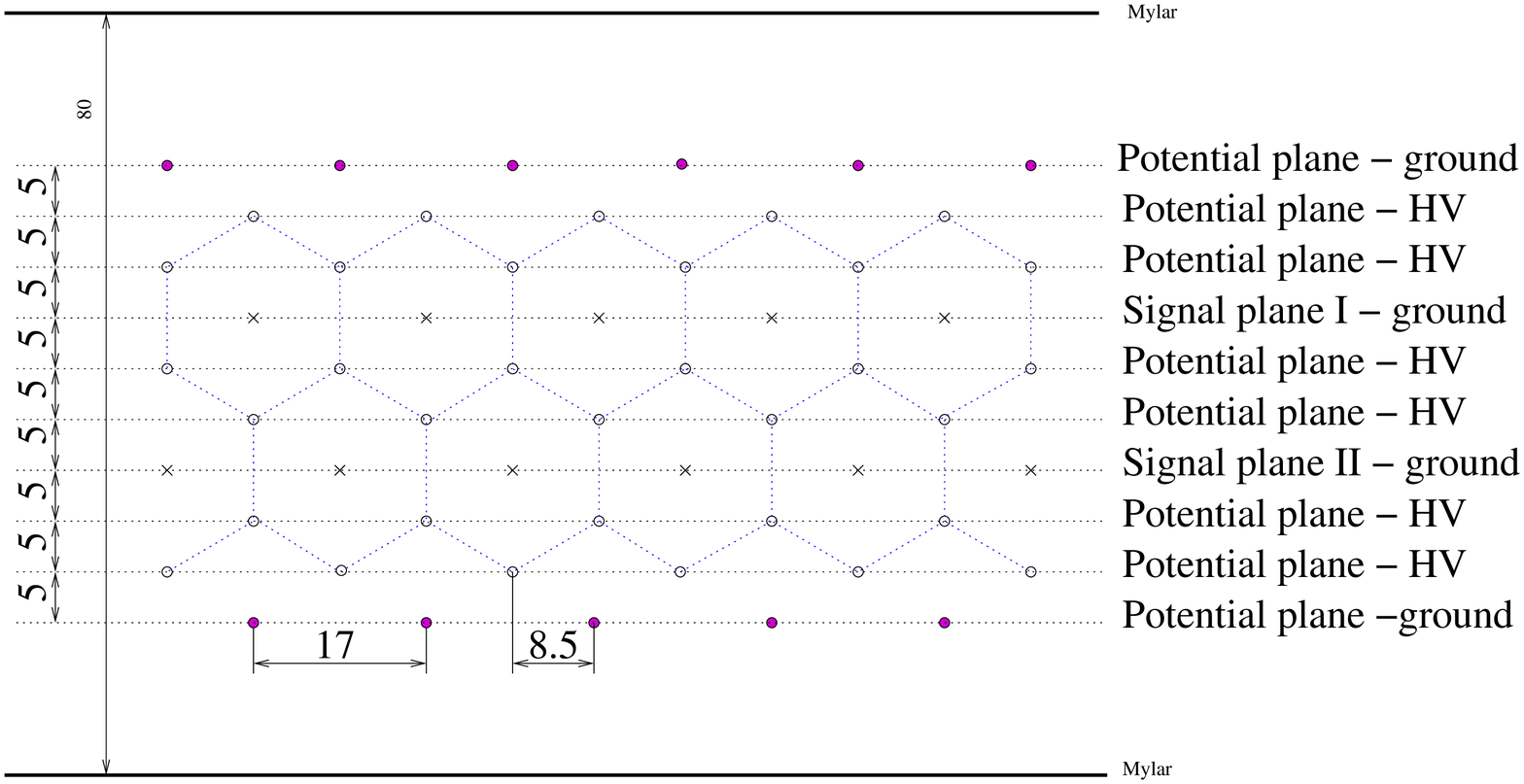} \\
    a) & b) \\
    \end{tabular}
    \caption{a) Horizontal distribution in the first focus of the accelerator 
    beam protons scattered over internal beryllium target.
    b) Drift chamber cross section. View along the wires}
    \label{fig:2}
\end{figure}

Each proportional chamber consists of two identical halves with mutually orthogonal sensitive wires,
providing {\it x} and {\it y} coordinates. The chambers have square sensitive region of $200\times200$~mm$^2$, 
1~mm signal wires pitch, 40~$\mu$m aluminum foil cathode electrodes and 6~mm between the cathodes.
We use "magic" gas mixture (argon-isobutane-freon) to feed the proportional chambers.
Beam tests showed efficiency better than 99\%. 
The chambers in the first focus measure the momentum of each pion
going to the target. Strong dipole magnets between the 
internal accelerator target and the first focus provide horizontal distribution of the
particles with different momentum with the dispersion of 57~mm/\%. A distribution 
over horizontal coordinate in the first focus is shown in fig.~\ref{fig:2}a for protons
scattered on the internal beryllium target. The peaks observed in the
picture correspond to (right to left) the elastic scattering, the first excitation of beryllium nucleus
and the second and the third excitations seen as one peak. This scattering was also used to get
an absolute calibration of the central beam momentum from the precise knowledge of the momentum of
the accelerator internal beam at three points 1057, 1195 and 1297 MeV/c. To ensure stability of the beam momentum 
an NMR monitoring of the magnetic field of the last dipole was used. This provided us with knowledge of
each incident particle momentum at the level of 0.1\%.

The liquid hydrogen target has a mylar cylinder container with diameter 40~mm and the length about 250~mm placed 
in high vacuum inside beryllium outer shell 1~mm thick. It is connected by two pipes 
to the liquefier system. One is used for liquid hydrogen inflow and through the other the evaporated gas gets back
to the liquefier. This design provides minimum of matter for the scattered particles. The refrigeration is provided by liquid
helium, which flow is controlled by the feedback supporting constant pressure of the hydrogen in the 
closed volume. This pressure corresponds to proper ratio between liquid and gas fractions of the hydrogen
and thus ensuring that the liquid occupies whole target volume and that the hydrogen is not frozen.
Pressures and temperatures in the target system are monitored and logged.
There are 8 one coordinate drift chambers in the elastic setup. 6 chambers have sensitive region 
$1200\times 800$~mm$^2$ and for 2 chambers closest to the target it is $600\times 400$~mm$^2$.
The chambers have double sensitive layers hexagonal structure 
shown in fig.~\ref{fig:2}b. Comparing to the conventional drift tubes this structure has much more 
complex fields, but provides significantly less amount of matter on the particle path. Potential wires form nearly 
regular hexagon with a side of 10~mm. Drift chambers are fed with 70\% Ar and 30\% CO$_2$ gas mixture. Beam
tests showed better than 99\% single layer efficiency and about 0.2~mm resolution for perpendicular tracks.

\begin{figure}[t!]
    \centering
    \includegraphics[width=0.95\textwidth]{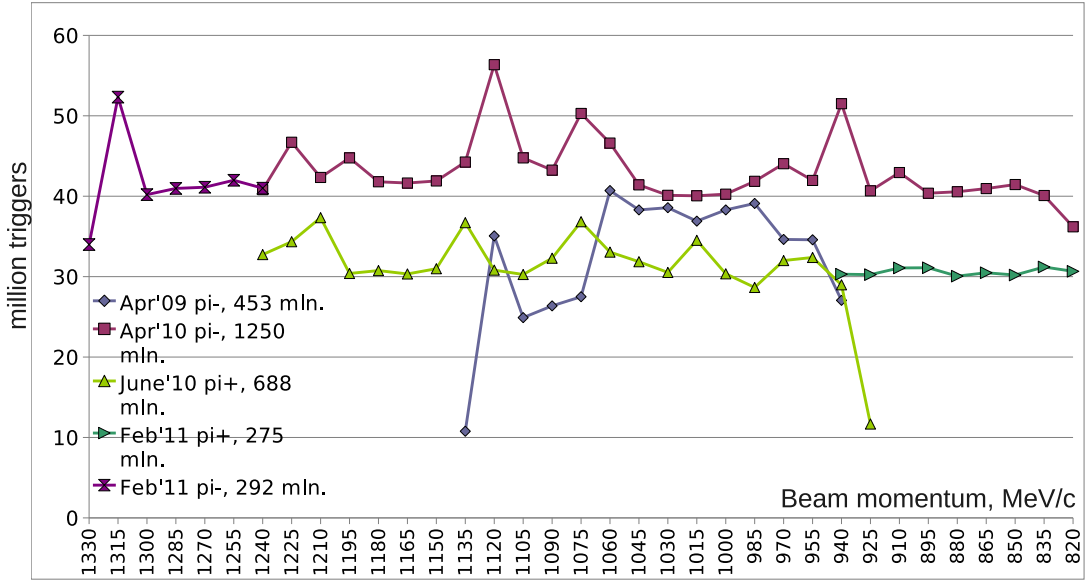}
    \caption{Distribution of the collected triggers (in millions) over runs and beam momentum intervals.}
    \label{fig:statistics}
\end{figure}

A unique distributed DAQ system based on the commercial 480~Mbit/s USB~2.0 interface was designed for the experiment~\cite{EPECURDAQ}. 
It consists of 100-channel boards for proportional chambers and 24-channel boards for drift chambers, 
placed on the chambers frames. Each board is connected by two cables (USB 2.0 and power) 
to the communication box, placed near the chamber. Then the data is transferred to the main 
DAQ computer by the standard TCP/IP connection. 
Trigger logic is capable of processing of several trigger conditions firing different sets of detectors. 
A soft trigger condition was used to acquire physics events:
$$
T = C_1 \cdot C_2 \cdot M_{1FCH} \cdot M_{2FCH} \cdot \overline{A_1}
$$
where $C_1$, $C_2$ and $A_1$ - signals from corresponding scintillation counters and
$M_{1FCH}$ and $M_{2FCH}$ - majority logic of the proportional chamber planes in the 1$^\mathrm{st}$ and
the 2$^\mathrm{nd}$ focuses. Other trigger conditions with large prescale were used for beam position and 
luminosity monitoring. A total statistics of nearly 3 billion triggers was collected. Figure~\ref{fig:statistics}
shows distributions of the collected triggers over runs and beam momentum intervals. 

\section{Data analysis}
\begin{figure}[t!]
    \centering
    \includegraphics[width=\textwidth]{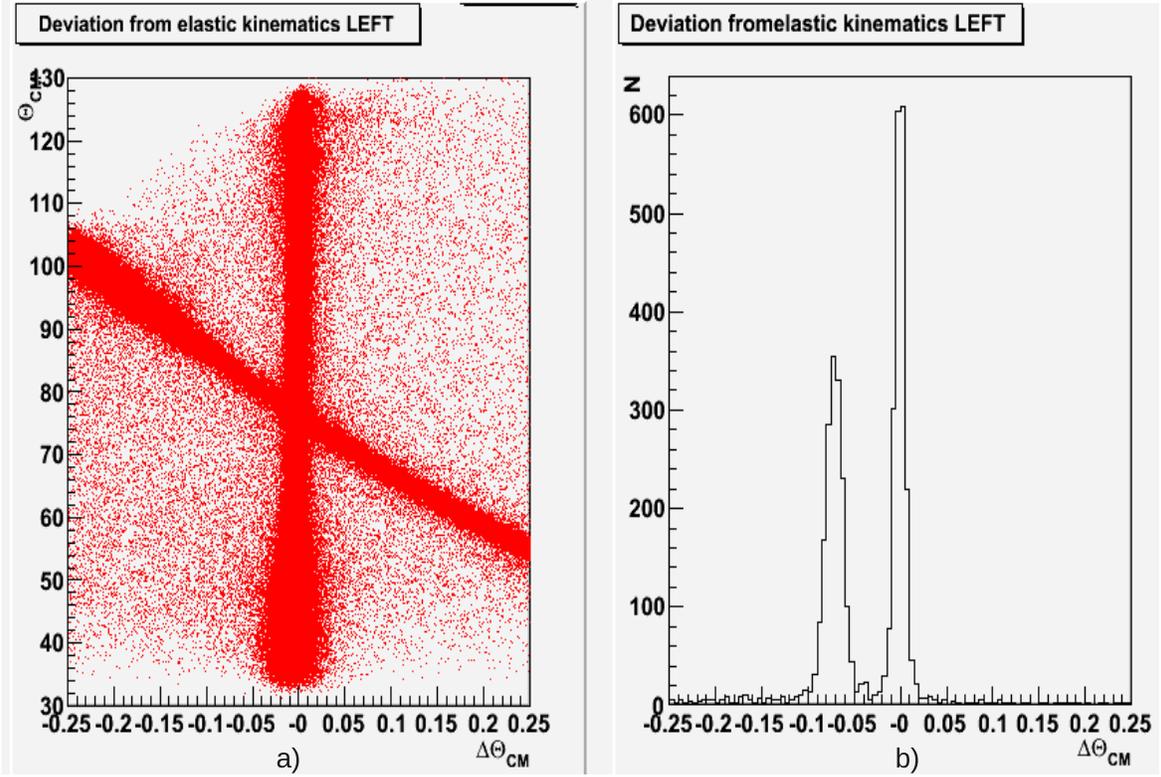}
    \caption{2-dimensional distribution of the events over reconstructed the difference between reconstructed
    scattering angles $\Delta\theta_{CM}$
    and the scattering angle $\theta_{CM}$ - (a) and its slice at $\theta_{CM} = 84^o$ (b).}
    \label{fig:QCM}
\end{figure}

Selection of the elastic events in this experiment is based on the angular correlation of pion 
and proton tracks. A single track is required in the beam chambers and the both scattering arms 
in each projection as well as that these tracks form a common vertex inside the target and lay in a plane.
A central of mass scattering angle $\theta_{CM}$ is calculated for both scattered particles under an assumption that
pion scattered to the left. A distribution of the events over the difference between reconstructed
scattering angles $\Delta\theta_{CM}$
and the scattering angle $\theta_{CM}$ is shown in fig.~\ref{fig:QCM}a for one beam momentum setting. 
Two clusters are clearly seen. One corresponds to the pion scattered to the left (the assumption was correct) and 
the other corresponds to the pion scattered to the right (the assumption was wrong). A slice of the
distribution for one degree $\theta_{CM}$ interval $\theta_{CM} = 84^o$ is shown in fig.~\ref{fig:QCM}b.

\begin{figure}[t!]
    \centering
    \begin{tabular}{ccc}
    \includegraphics[width=0.31\textwidth]{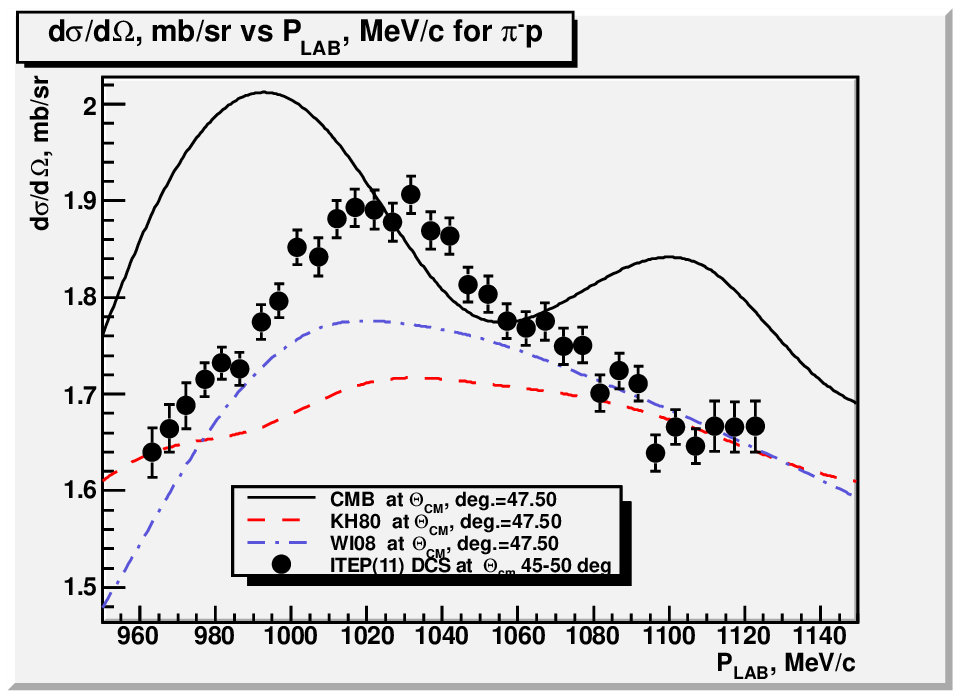} &
    \includegraphics[width=0.31\textwidth]{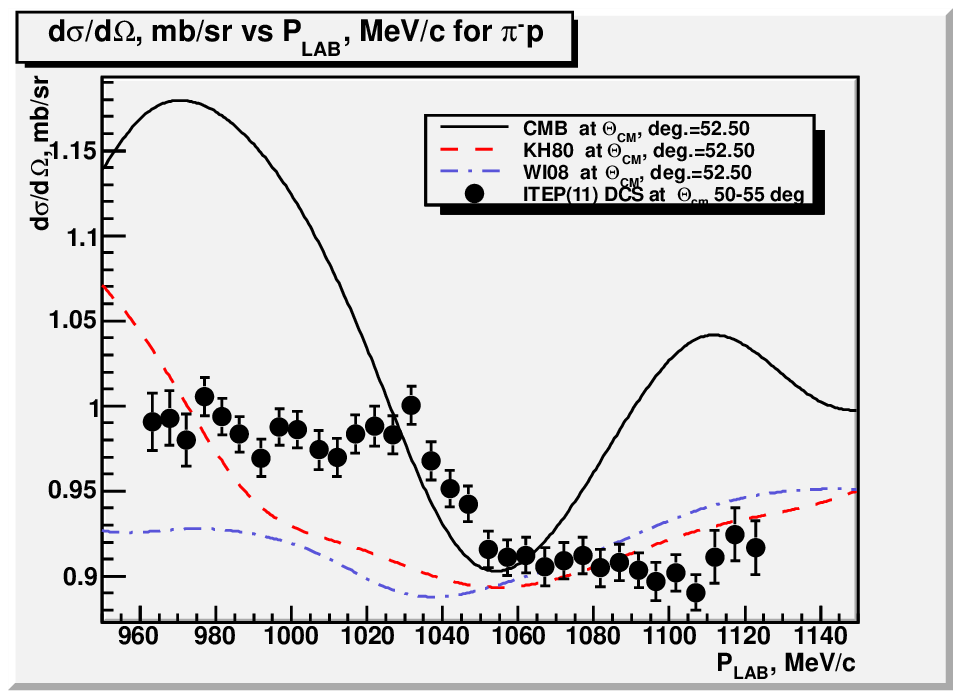} & 
    \includegraphics[width=0.31\textwidth]{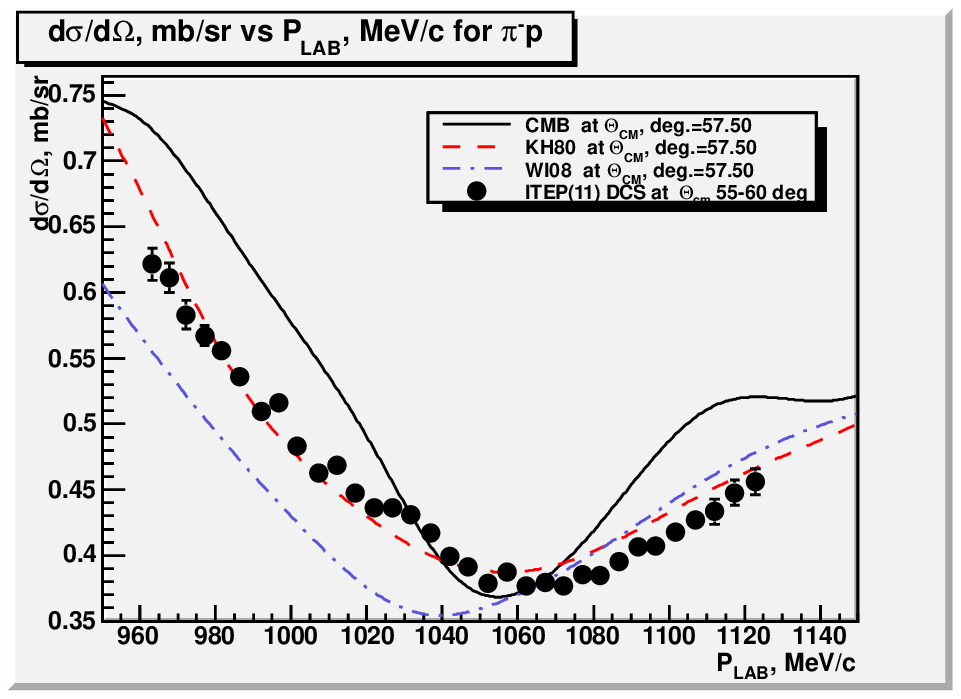} \\
    \includegraphics[width=0.31\textwidth]{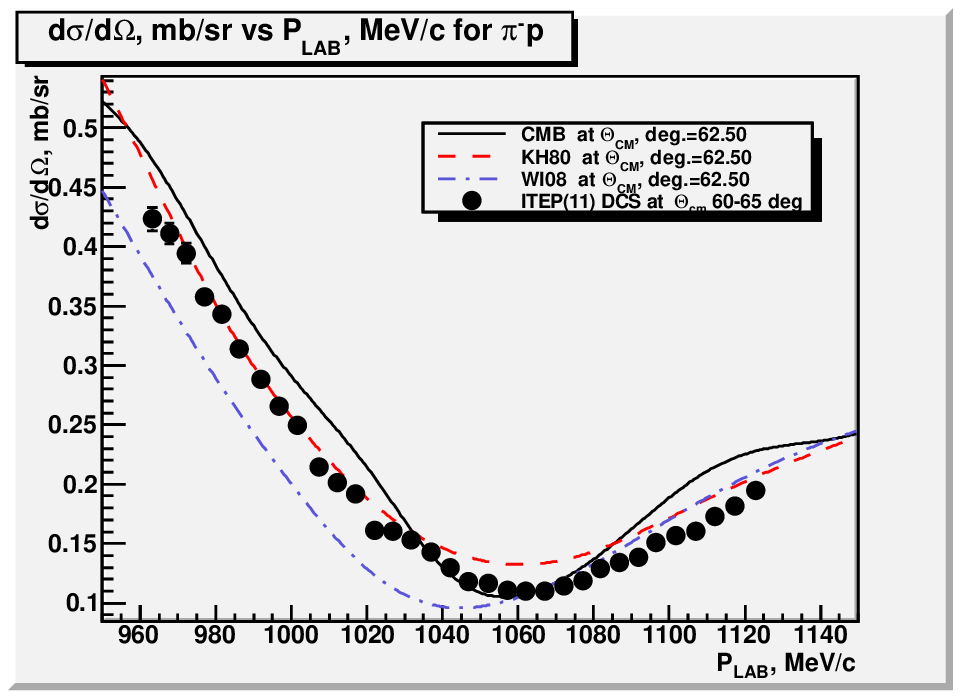} &
    \includegraphics[width=0.31\textwidth]{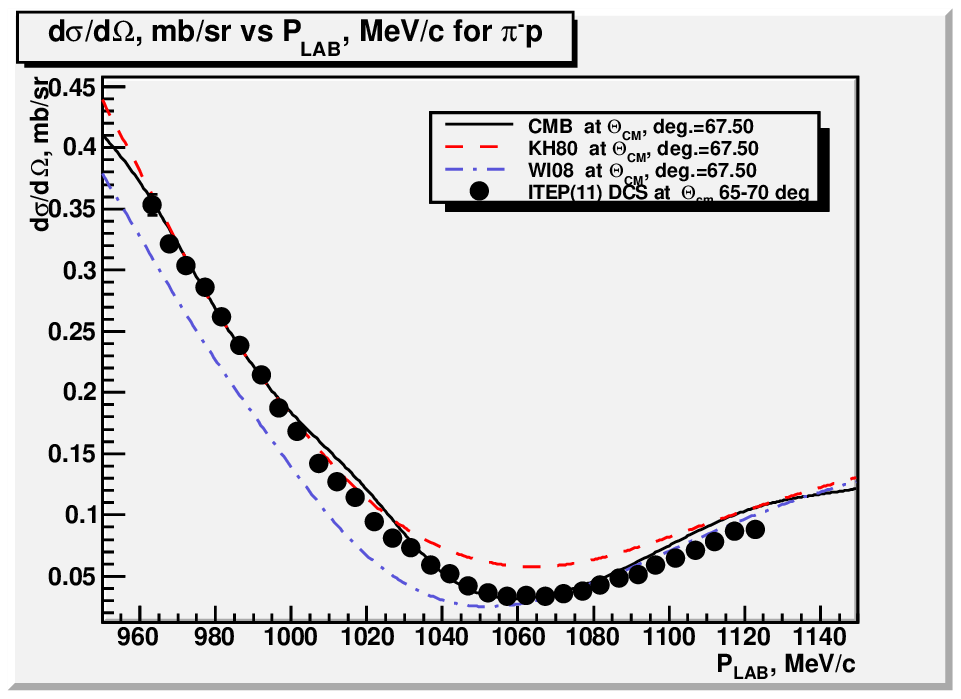} &
    \includegraphics[width=0.31\textwidth]{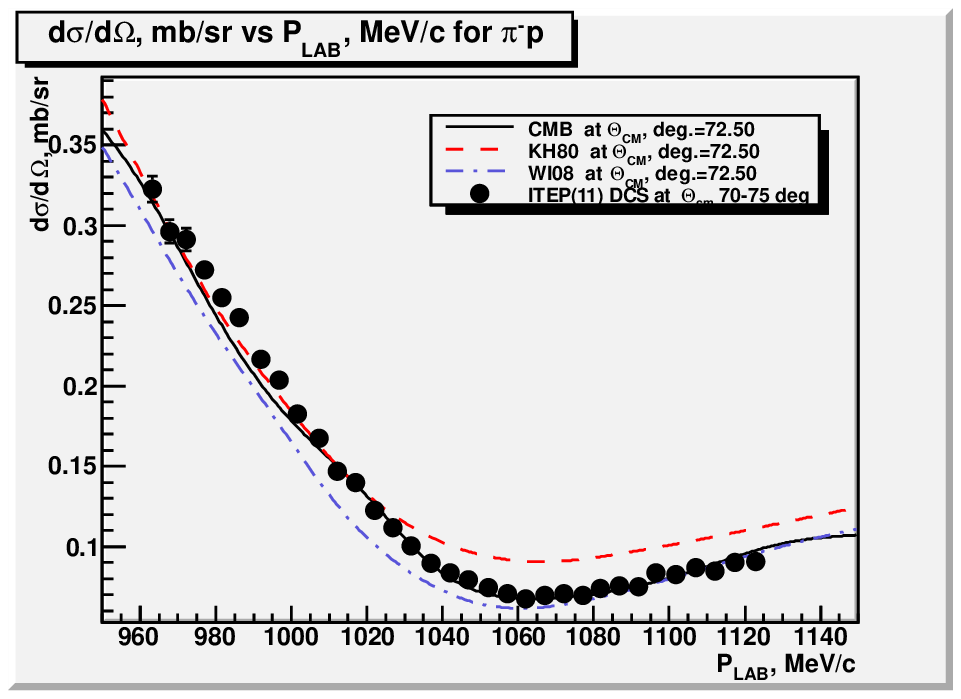} \\
    \includegraphics[width=0.31\textwidth]{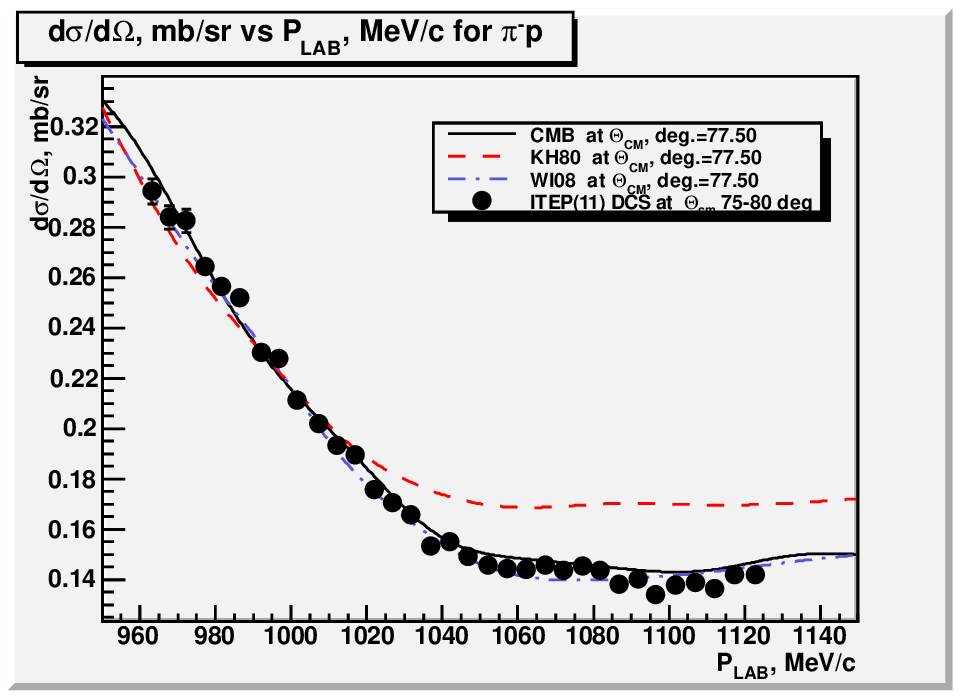} &
    \includegraphics[width=0.31\textwidth]{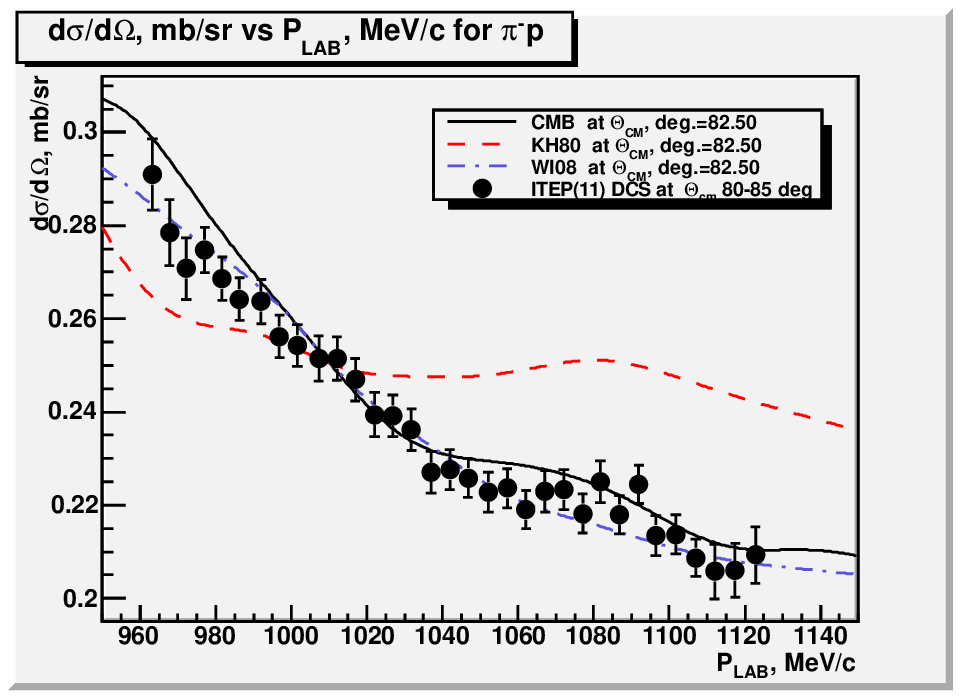} &
    \includegraphics[width=0.31\textwidth]{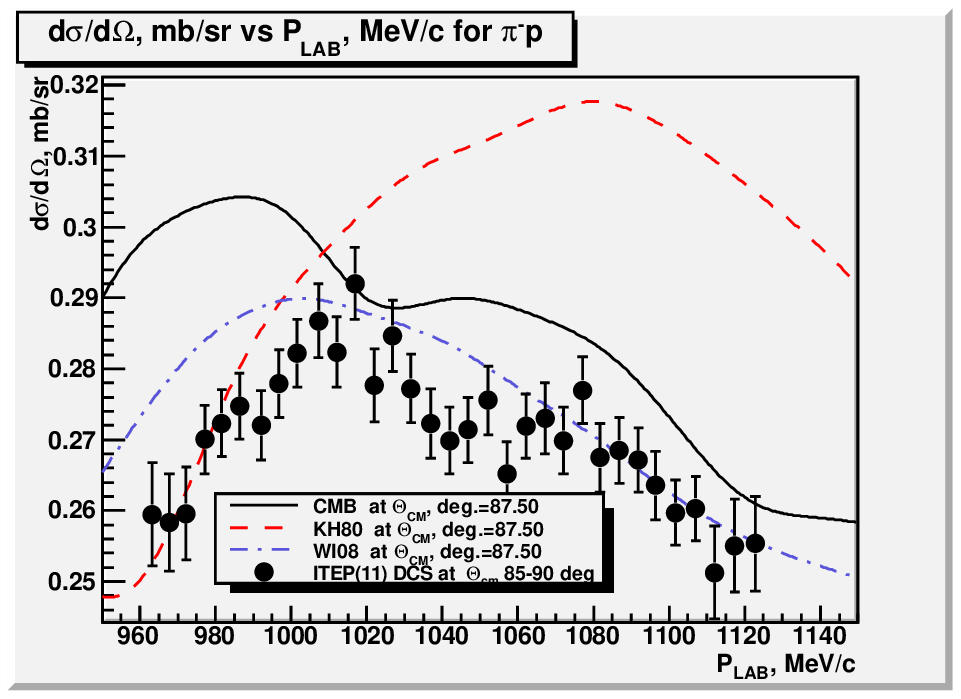} \\
    \includegraphics[width=0.31\textwidth]{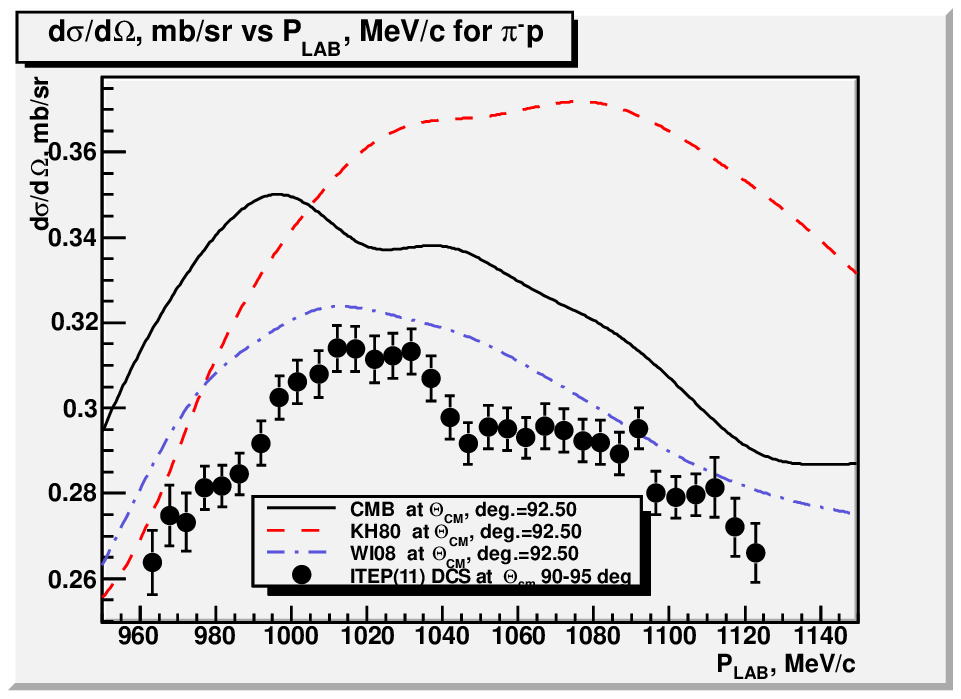} &
    \includegraphics[width=0.31\textwidth]{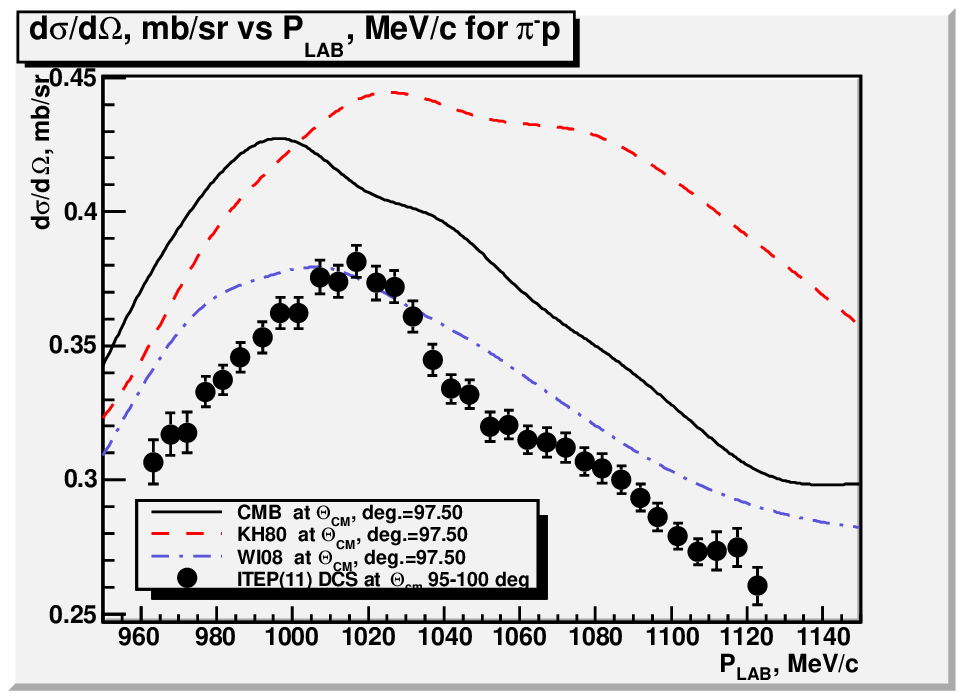} &
    \includegraphics[width=0.31\textwidth]{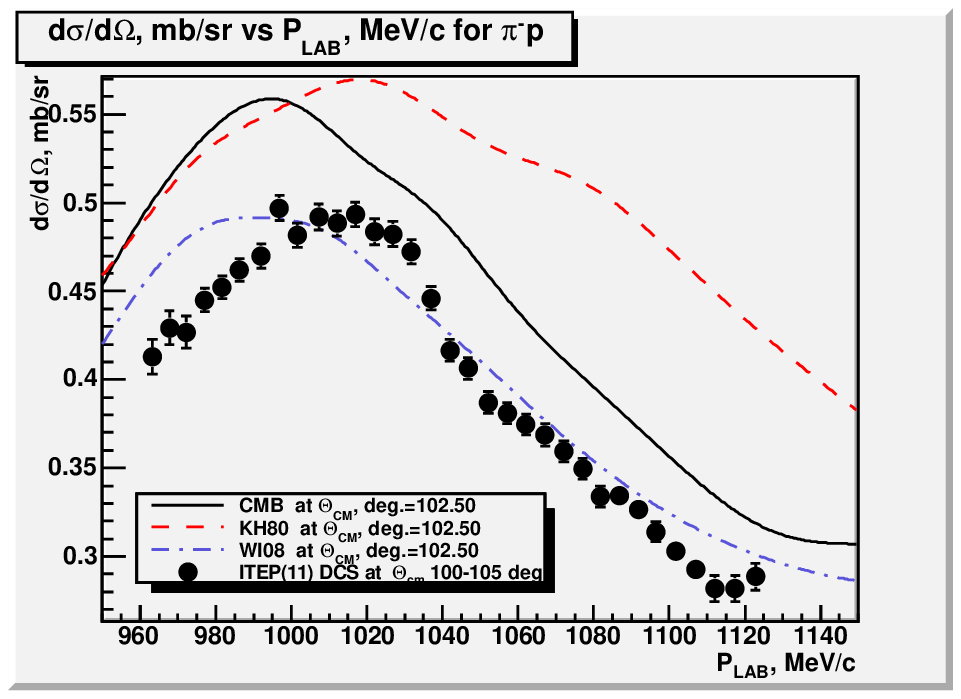} \\
    \includegraphics[width=0.31\textwidth]{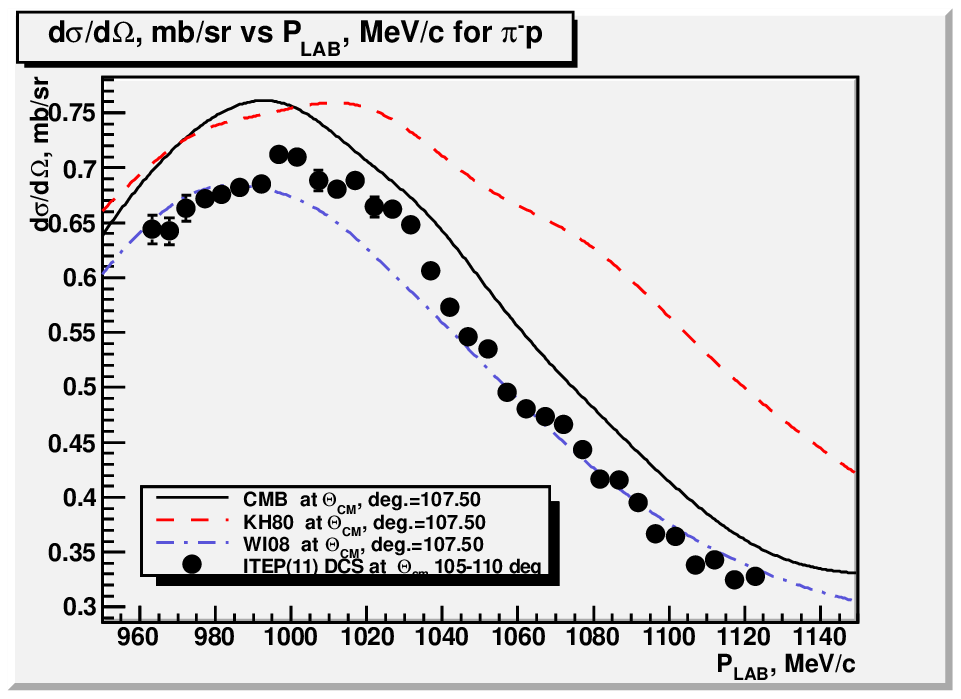} &
    \includegraphics[width=0.31\textwidth]{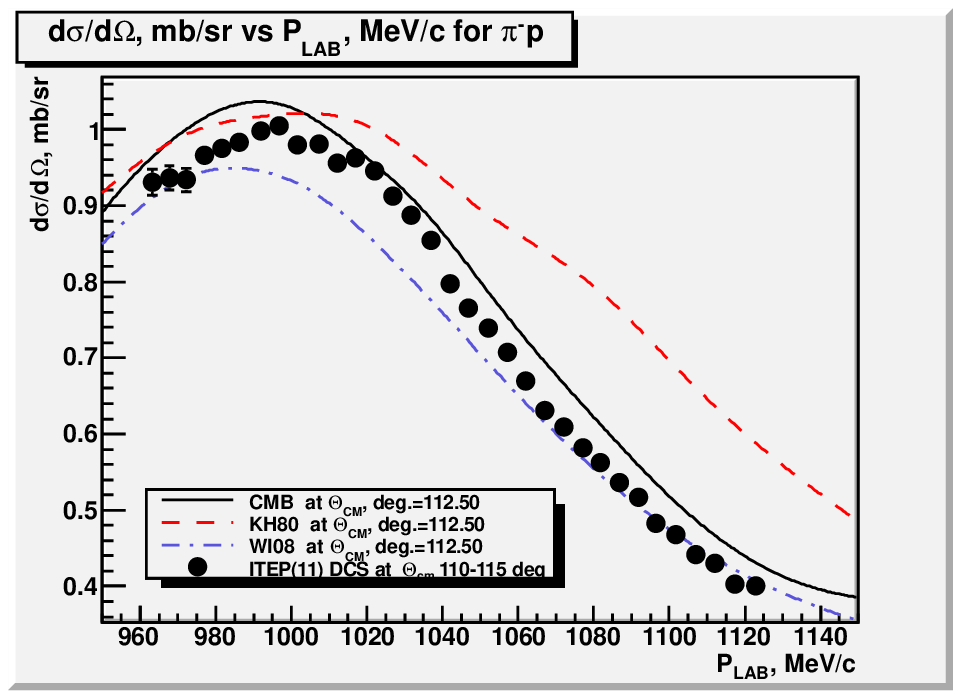} &
    \includegraphics[width=0.31\textwidth]{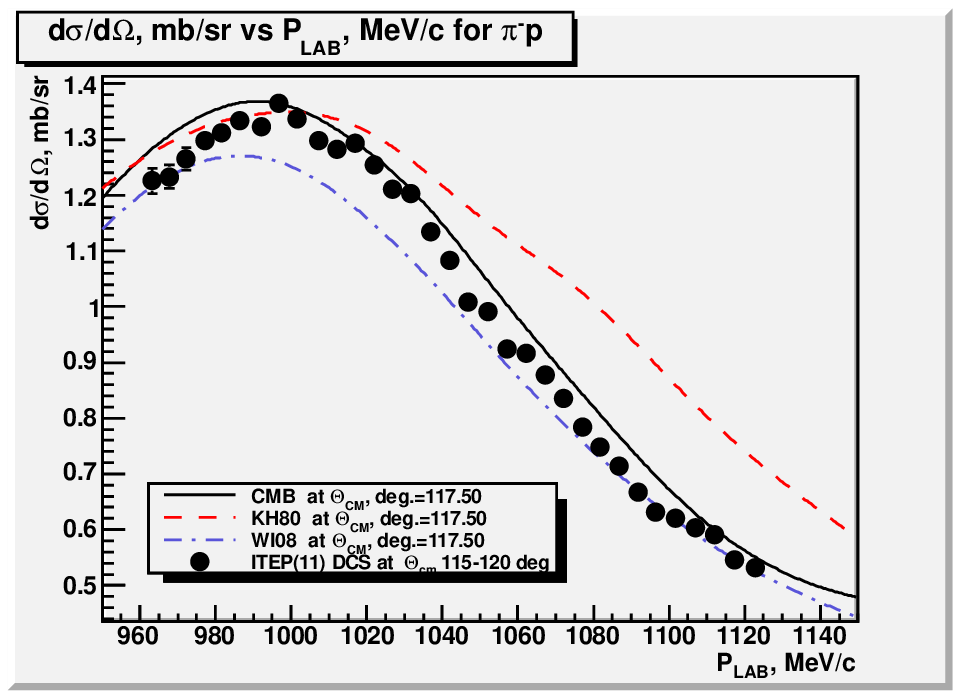} \\
    \end{tabular}
    \caption{Differential cross section of the elastic $\pi^-p$-scattering as a function of the 
    incident pion momentum for 5-degree $\theta_{CM}$ intervals. {\bf Preliminary.}}
    \label{fig:dcs}
\end{figure}

Preliminary results on the differential cross section of $\pi^-p$ elastic scattering
are shown in fig.~\ref{fig:dcs} as
a function of the incident pion momentum for 5-degree $\theta_{CM}$ intervals. These data
corresponds to the analysis of about 15\% of the statistics already collected. Some irregularity 
is manifested at $p_{beam} = 1020-1040$~MeV/c. The nature of the irregularity could be either
connected to a narrow resonance with mass around 1690~MeV and width of 5-10~MeV 
or to a threshold effect ~\cite{BAZ}, caused by opening of
the channels $\pi^-p \to K^0\Sigma^0$ ($\sqrt{s}=1690.2$~MeV) and $\pi^-p \to K^+\Sigma^-$
($\sqrt{s}=1691.1$~MeV). The resonance, if it is a memeber of the pentaquark antidecouplet,
should be in P-wave, while the threshold effect should manifest itself in S-wave.
The structure observed in the differential cross section is a result of an interference of a fast 
change in some partial wave with slow changing non-resonant backgound.
We plan to collect large statistics in a narrow region $p_{beam} = 1000-1070$~MeV/c, 
which will allow us to plot data in more fine angle and energy binning in order to
find out which wave is affected. Further analysis of the data already collected,
including data collected with positive pions, is also under way.

\section*{Acknowledgments}
The experiment is supported by the Russian State Atomic Energy Corporation ROSATOM and by Russian 
Fund for Basic Research grants 12-02-00981-а, 09-02-00998a and 05-02-17005a.

\end{document}